\documentclass[12 pt]{article} 
\usepackage{ucs} 
\usepackage[utf8x]{inputenc} 
\usepackage[english]{babel}  

\usepackage{amsmath, amsfonts, amssymb}
\usepackage{graphicx}
\usepackage{indentfirst}
\usepackage{threeparttable}
\usepackage{url}

\def\la{\;
\raise0.3ex\hbox{$<$\kern-0.75em\raise-1.1ex\hbox{$\sim$}}\; }
\def\ga{\;
\raise0.3ex\hbox{$>$\kern-0.75em\raise-1.1ex\hbox{$\sim$}}\; }

\begin{document}

\title{\LARGE \bf Indication of the electron-to-proton mass ratio variation within the Galaxy}

\maketitle

\begin{center}
J. S. Vorotyntseva\footnote{j.s.vorotyntseva@mail.ioffe.ru}, S. A. Levshakov\\
{\it Ioffe institute, Politechnicheskaya 26, St. Petersburg 194021 Russia}
\end{center}

\bigskip
\begin {abstract}
Near ($\approx 100$ pc) and far ($\approx 8.7$ kpc)  
relative to the Galactic center, the molecular clouds SgrB2(N) and
Orion-KL exhibit different values of the fundamental 
physical constant $\mu=m_{\rm e}/m_{\rm p}$ -- the  electron-to-proton mass ratio.
Measured frequency difference between the emission lines of methanol (CH$_3$OH), 
-- $J_{K_u} \to J_{K_\ell} = 6_3 \to 5_2 A^+$ 542000.981 MHz, $6_3 \to 5_2 A^-$ 542081.936 MHz,
and $8_0 \to 7_{-1} E$ 543076.194 MHz, --
 observed with the space observatory {\it Herschel}
toward SgrB2(N) and Orion-KL corresponds to (Sgr-Ori): 
$\Delta\mu/\mu = (-3.7 \pm 0.5)\times10^{-7}$ (1$\sigma$ C.L.).
At the same time, comparison of the same methanol lines in Orion-KL with laboratory frequencies shows 
no significant changes in $\mu$ (Ori-lab): $\Delta\mu/\mu = (-0.5 \pm 0. 6)\times10^{-7}$, while a comparison between SgrB2(N) and laboratory lines indicates a lower value of $\mu$ near the Galactic center (Sgr-lab): 
$\Delta\mu/\mu = (-4.2 \pm 0.7)\times10^{-7}$.
 The reduced value of $\mu$ in SgrB2(N)
is not explained by known systematic effects and requires further investigation.
\end{abstract}

\section{Introduction}

Precision measurements of fundamental physical constants remain a
longstanding challenge for both laboratory and space research.
Under terrestrial conditions, experiments with atomic clocks, and, more recently, with nuclear clocks, have achieved 
extremely tight constraints on changes with time of dimensionless physical constant~-- the fine
structure constant ($\alpha=e^2/\hbar c$) at the level of 10$^{-19}$~ yr$^{-1}$,
and the electron-to-proton mass ratio
($\mu$=$m_{\rm e}/m_{\rm p}$) at the level of 10$^{-16}$~yr$^{-1}$.
Differential measurements of these quantities at different points of the Earth's orbit under conditions of changing
gravitational potential of the Sun also did not reveal appreciable deviations in the frequencies
of the atomic clocks at the level of 10$^{-8}$. Extensive material on this topic was collected by
and analyzed in a recent review \cite{Uz}.

Spatial and temporal changes in physical constants are predicted in various theories extend 
the Standard Model of particle physics. The necessity for such
 extension arises, in particular, from attempts to explain the nature of dark matter.
Since it has not been possible to detect dark matter particles so far,
suggestions have been made about scalar fields that could manifest  themselves as dark matter. These fields
can modulate the masses of elementary particles, such as electrons and quarks,  
and, hence, should lead to small variations of $\mu$ \cite{On}-\cite{SF}.
Changes of this kind can affect in turn the structure of energy levels in molecules,
which can be detected experimentally.

The search for spatial and temporal variations in $\mu$ is related to estimating the quantity
 \begin{equation}
\Delta\mu/\mu = (\mu_{obs}-\mu_{lab})/\mu_{lab},
\label{Eq1}
\end{equation}
where $\mu_{lab}$ -- the laboratory value of $\mu$, and
$\mu_{obs}$ -- the value observed in the astronomical objects.

This comparison is possible due to the fact that the electro-vibro-rotational transitions
in molecular spectra have a specific dependence on the local value of $\mu$,
which is individual for each transition \cite{VL, KL}.  
The dependence on $\mu$ is characterized by a dimensionless sensitivity coefficient, $Q$, which shows the reaction
of a given molecular transition with a frequency $f$ to a small change in $\mu$:
 \begin{equation}
Q=\frac{df/f}{d\mu/\mu},
\label{Eq2}
\end{equation}
where $df/f$ is the relative frequency shift, and $d\mu/\mu$ is determined by (\ref{Eq1}). 
The values of $Q$ are calculated using quantum-mechanical methods for the transition $f=E_{u}-E_{\ell}$ 
using the so-called $q$-factor \cite{LKR}:
 \begin{equation}
\Delta f=q\frac{\Delta\mu}{\mu}\,  ,
\label{Eq3}
\end{equation}
where $\Delta f$ is the frequency change caused by a small variation of $\mu$,
$q = q_{u}-q_{\ell}$ ( in cm$^{-1}$), and
 \begin{equation}
Q=\frac{q}{f}.
\label{Eq4}
\end{equation}

To estimate hypothetical variations of $\mu$, it is necessary to use pairs of lines
with different sensitivity coefficients $Q_1$ and $Q_2$ observed in the same
molecular cloud:
 \begin{equation}
\frac{\Delta\mu}{\mu} =\frac{V_1-V_2}{c(Q_{2}-Q_{1})},
\label{Eq5}
\end{equation}
where $V_1$ and $V_2$ are the measured radial velocities, and $c$ is the speed of light.
Note that the conversion to the velocity scale from the frequency scale is carried out according
to the radio astronomical convention
 \begin{equation}
\frac{V}{c} = \frac{f_{lab} −f_{sky}}{f_{lab}}.
\label{Eq6}
\end{equation}

At present, the most stringent constraints on $\mu$-variations in the Galactic disk 
 were obtained from observations of the inversion transition of ammonia NH$_3$
 ($Q$ = 4.46~\cite{FK}) and purely
rotational transitions ($Q$=1) HC$_3$N, HC$_5$N and HC$_7$N:
$\Delta\mu/\mu < 7 \times 10^{-9}$ \cite{LHR}.
Another perspective molecule for such purposes turned out to be methanol CH$_3$OH, and its isotopologues
-- methanol transitions have high sensitivity coefficients of both signs \cite{LKR, JXK, VKL}. 
Constraints on $\mu$-variations based on observations of methanol in the Galaxy
are established at the level of $\Delta\mu/\mu < 1\times10^{-8}$ \cite{VL24},
and for extragalactic objects (at redshift $z$=0.89) $\Delta\mu/\mu < 5\times10^{-8}$ \cite{KUM}.

Thus, the expected
effect of hypothetical scalar fields on the masses of elementary particles was not revealed
at this accuracy level.

\section{Methods}
The most significant problem in achieving a higher accuracy level ($\sim~10^{-9}$)
is the errors in laboratory frequencies, which are for methanol lines
reach values of 50 kHz in the high frequency range ($f\sim 540$ GHz).
To reach the level of $10^{-9}$,
the accuracy must be on the order of several kHz, which is problematic in
laboratory measurements. 
At the same time, astronomical observations make it possible to achieve
a rather high accuracy acceptable for precision measurements of
$\Delta\mu/\mu$ at the level of $10^{-9}$.

To minimize the dependence on laboratory frequencies, a slightly modified approach can be used to estimate the $\mu$-variations, which is as follows.
We offer to compare the difference of the measured frequencies ($f_1 - f_2$) of the same pairs
of methanol lines with different sensitivity coefficients $Q_1$ and $Q_2$
observed in two astronomical objects ($a$) and ($b$) at different galactocentric distances. It follows from the relations (\ref{Eq1}-\ref{Eq4}) that
 \begin{equation}
\frac{\Delta\mu_{a,b}}{\mu} = \frac{\mu_a - \mu_b}{\mu_{lab}} =
\frac{(f_{1,a}-f_{2,a}) - (f_{1,b}-f_{2,b})}{\Delta q},
\label{Eq7}
\end{equation}
where $f_{1,a}, f_{2,a}$ and $f_{1,b}, f_{2,b}$ -- the observed transition frequencies in the comoving
reference frame (corrected for the radial velocity $V_{LSR}$ of the object),
and $\Delta q = q_1 - q_2$ -- the difference of the $q$-factors of the compared lines. 
This method allows us to use only astronomical frequencies without resorting to laboratory measurements.

\section{Observations and parameters of {\it Herschel} spectra}
The spectra used in this work were taken from the archive
of the space observatory {\it Herschel}\footnote{{\it Herschel} is an ESA space observatory 
with science instruments provided by European-led Principal Investigator consortia and with
important participation from NASA.} \cite{Hersch}. 
The observations were obtained using
the instrument{\it HIFI} (Heterodyne Instrument for the Far-Infrared)
with the Wide-Band Spectrometer {\it WBS} in {\it Spectral Scan} mode. 
This mode included
observations with frequency tuning of the local oscillator (LO), which made it possible to obtain, after separating the lower ({\it LSB}) and upper ({\it USB}) frequency bands, a single spectrum of the object in a wide spectral range.

The technical characteristics of the observed range
of 480--560 GHz (band 1$a$) are as follows: the telescope's aperture at the center of the range is $40^{\prime\prime}$,
the width of the intermediate frequency band (IF) is 4 GHz, and the spectral resolution (channel width) 
$\Delta_{ch}\sim$ 0.3 km/s (500 kHz).

The median calibration error of the frequency scale for each of the four CCD arrays
 of {\it WBS} does not exceed 50 kHz with almost zero average shift of the center frequency of the spectral line for various LO frequencies in the test mode \cite{Av}. 
Also, no trends were detected within the intermediate frequency band in the CCD pixel-to-frequency transform.

From all the astronomical objects observed by the {\it Herschel}, we selected two~-- Orion-KL and SgrB2(N).
Orion-KL is a large molecular gas complex in Orion, where star formation is actively taking place.
SgrB2(N) is a massive molecular gas cloud in Sagittarius,
located near the Galactic center, in which $\Delta\mu/\mu$ estimates have never been performed before.
The object coordinates~--
right ascension ($R.A.$) and declination ($Dec.$), as well as the distance from
the Galactic center ($D_{GC}$) are shown in  Table~\ref{T1}. 

\begin{table*}
\centering
\caption{ Object coordinates \cite{Hersch} and galactocentric distances $D_{GC}$}
\label{T1}
\begin{tabular}{l c c c}
\hline
{\bf{Object}}&$R.A.$ (J2000) &$Dec.$ (J2000) & $D_{GC}$, kpc\\
\hline
{\bf{Orion-KL}}&05$^h$35$^m$14.36$^s$ &--05$^\circ 22^\prime 33.63^{\prime\prime}$ & 8.7\\
{\bf{SgrB2(N)}}&17$^h$47$^m$20.06$^s$ &--28$^\circ 22^\prime 18.33^{\prime\prime}$ & 0.1\\
\hline

\end{tabular}
\end{table*}

We used high-level spectra (level 2.5) from the {\it{Herschel}} archive.
The processing level 2.5 implies conversion to the scales of antenna temperatures $T^\ast_A$ (in Kelvin)
and observed frequencies $f_{sky}$ (in GHz) \cite{TAB}. The spectra for each astronomical object
are provided in two polarizations --
horizontal $H$ and vertical $V$.
We subtracted the continuum from each 
of these spectra and stacked them with
weights inversely proportional to the squares of the noise levels.

\section{Calculated line parameters}
It is known from {\it{Herschel}} spectral surveys that
microwave spectra of Orion-KL and SgrB2(N) \cite{CBN, NBL}
 have a rich molecular composition especially in the frequency band
 480--560 GHz~-- from simple molecules
(e.g., CO), to complex organic ones such as methanol (CH$_3$OH).
For our purposes, we selected three close methanol lines 
 in the Orion-KL and SgrB2(N) spectra that fall
in the same intermediate frequency band IF = 4 GHz -- two lines at 542 GHz and one at 543 GHz.
This selection minimizes possible instrumental systematic errors, associated with {\it{WBS}} calibrations.
The parameters of the selected methanol lines are presented in Table~\ref{T2}: transition,
which can be described by two quantum numbers~-- the total angular momentum, $J$, and 
its projection onto the symmetry axis of the molecule, $K$, for the upper $u$ and lower $\ell$ levels;
 the laboratory frequency $f_{lab}$; the line strength
$S\mu_e^2$;  the energy of the lower level $E_{\ell}$ \cite{XL},
and the sensitivity coefficient $Q$ \cite{JXKU}.

\begin{table*}
\centering
\caption{Parameters of the selected CH$_3$OH transitions. Given in parentheses are 
the
laboratory errors in the last digits.}
\label{T2}
\begin{tabular}{l c c c c c }
\hline
\multicolumn{1}{l}{Transition} &$f_{lab}$,& $S\mu_e^2$,& $E_{\ell}$,& $Q$&\\
$J_{K_u} \to J_{K_\ell}$ &MHz&$D^2$&cm$^{-1}$&\\
\hline
{$6_{3} \to 5_{2}A^+$}&542000.981(50) &5.5658&50.413&0.0\\   
{$6_{3} \to 5_{2}A^-$}& 542081.936(50)&5.5638&50.411&0.0\\ 
{$8_{0} \to 7_{-1}E$}& 543076.194(50) &4.1354&49.035&1.7\\ 
\hline
\end{tabular}
\end{table*}

Table~\ref{T2} shows that the molecular
transitions have close energies of the lower levels, which means that they emit from the same region 
of the molecular cloud.

\begin{figure}
\vspace{-6.0cm}
\centering
\includegraphics[width=1.0\textwidth]{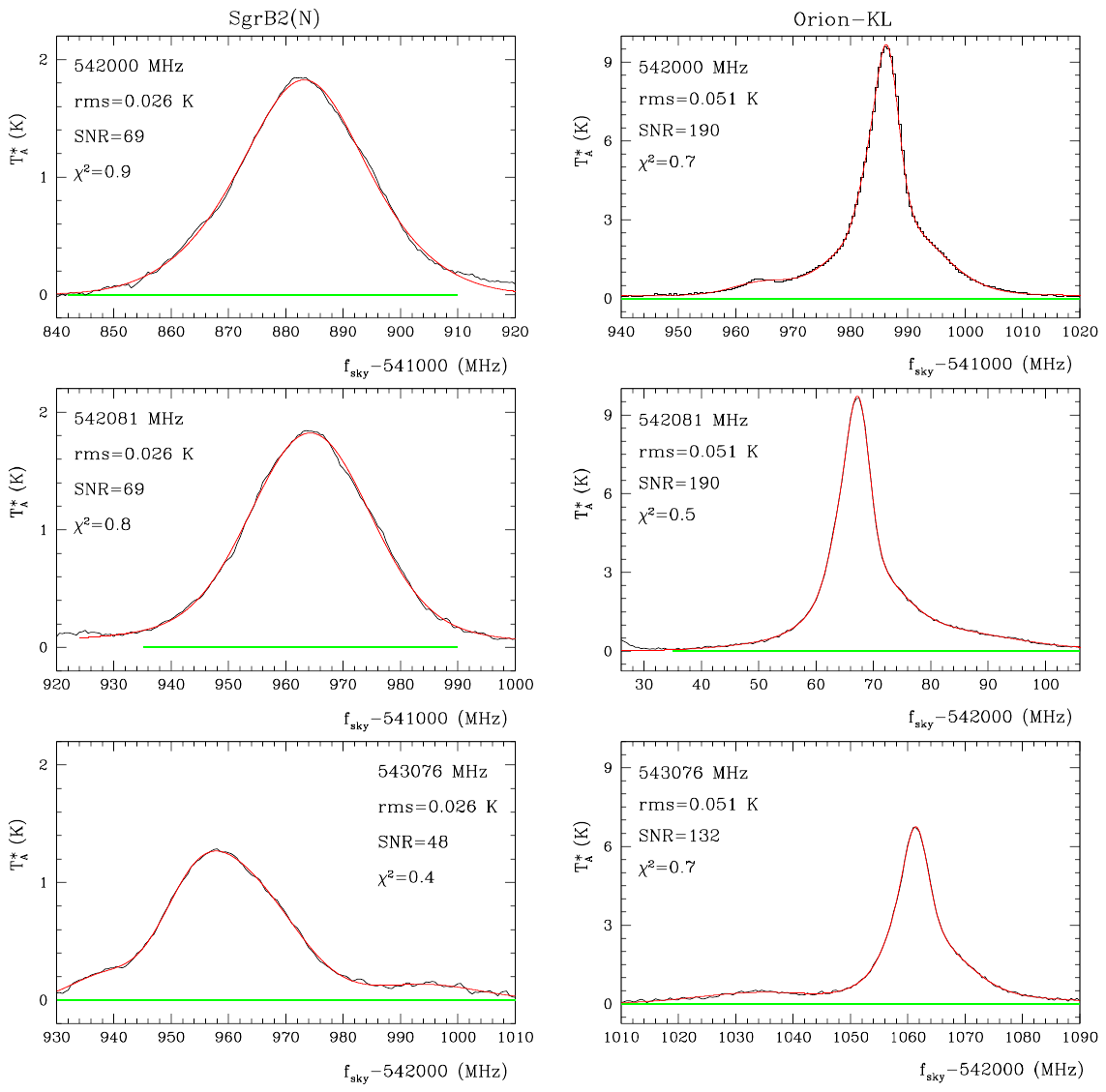}
\vspace{-3.5cm}
\caption{\small Methanol lines toward the two sources -- SgrB2(N) and Orion-KL, obtained
at the Herschel Space Observatory. The original spectra are shown in black, and the model spectra are shown in red. The horizontal green lines mark the ranges used in minimizing $\chi^2$.
The line parameters are listed in Table ~\ref{T3}.
}
\label{F1}
\end{figure}

We used a multi-component Gaussian-like model to describe the line profile.
At the same time, the minimum number of Gaussian
components was decided, which provides $\chi^2_\nu\la 1$.
The resulting envelope curves are shown in Fig.~\ref{F1} in red color: left panel~-- the 
observed lines toward  SgrB2(N), right panel~-- toward Orion-KL.
For each line, the laboratory frequency, the noise level $rms$, the
signal-to-noise ratio $SNR$, and the value of $\chi^2_\nu$ are specified.
The parameters of the observed lines are listed
in Table~\ref{T3} -- the frequency at the envelope curve maximum $f_{sky}$, the corresponding radial velocity $V_{LSR}$, and 
the full width at half maximum line width $FWHM$.

Also, Table~\ref{T3} lists the observed frequencies in the comoving reference frame $f_c$ 
corrected for the average radial velocity $V_{LSR}$. It is determined
along the lines $6_{3}-5_{2}A^+$ and $6_{3}-5_{2}A^-$, which have zero sensitivity coefficients $Q$.
For SgrB2(N), the average velocity $\langle V_{a} \rangle$ = 65.12(2) km~s$^{-1}$,
for Orion-KL -- $\langle V_{b} \rangle$ = 8.15(2) km~s$^{-1}$.

The last column in Table~\ref{T3} gives an estimate of the expected accuracy of the line position.
The expected statistical error of the center of a single Gaussian line profile is defined by the expression~\cite{LRT}
 \begin{equation}
\sigma_f = 0.7 \frac{\Delta_{ch}}{SNR}\sqrt{FWHM/\Delta_{ch}}.
\label{Eq8}
\end{equation}
Table~\ref{T3} shows  that the analytical error estimates from (\ref{Eq8}) are almost identical to the numerical ones given in parentheses in the second column, which were obtained by the $\Delta \chi^2$ method applied to the envelope curves.

\begin{table*}
\centering
\caption{Parameters of the observed lines. The statistical errors in the last digits are shown in parentheses. (1$\sigma$).}
\label{T3}
\begin{tabular}{l c c c c c }
\hline
\multicolumn{1}{l}{\bf{Object}} &$f_{sky}$,& $V_{LSR}$,& $FWHM$,&$f_c$,&$\sigma_f$,\\
 &MHz&km~s$^{-1}$&MHz&MHz&kHz\\
\hline
{\bf{Orion-KL}}&541986.222(8) &8.16(3)&7.3&542000.964(8)&7\\   
&542067.209(8) &8.15(3)&7.2&542081.953(8)&7\\   
&543061.376(12)&8.18(3)&7.7&543076.147(12)&10\\   
&&&&&\\
{\bf{SgrB2(N)}}&541883.185(38)&65.16(4)&26.1&542000.921(38)&37\\   
&541964.243(37)&65.09(3)&25.6&542081.996(37)&36\\   
&542957.837(51) &65.34(4)&23.7&543075.806(51)&50\\   

\hline
\end{tabular}
\end{table*}

\section{$\Delta\mu/\mu$ estimations and discussion}
As noted above, to estimate the value of $\Delta\mu_{a,b}/\mu$
two massive molecular clouds were chosen -- SgrB2(N) (object {\it{a}}) and Orion-KL (object {\it{b}}).
Estimates on $\mu$-variations are obtained from comparing the differences in the measured frequencies
of methanol lines converted into the comoving reference frames of the {\it a} and {\it b} objects.
Transitions with $Q = 0$ at 542000 MHz and
542081 MHz, which were compared with the 543076 MHz transition having $Q$ = 1.7,
led to the following results:
 \begin{equation}
\frac{\Delta\mu_{a,b}}{\mu} = (-3.2\pm0.7)\times10^{-7}
\label{Eq9}
\end{equation}
for the 543076 and 542000 MHz lines, and
 \begin{equation}
\frac{\Delta\mu_{a,b}}{\mu} = (-4.2\pm0.7)\times10^{-7}
\label{Eq10}
\end{equation}
for the 543076 and 542081 MHz lines.\\
This provides the average value
 \begin{equation}
\langle {\Delta\mu_{a,b}}/{\mu} \rangle = (-3.7\pm0.5)\times10^{-7}.
\label{Eq11}
\end{equation}

The resulting error $\sigma_{\Delta\mu/\mu}$ in this case is calculated
 as the square root of the sum of the squares
of the measured frequency errors (see the fifth column in Table~\ref{T3}) divided by the difference of the $q$-factors
$\Delta q$ (this value is assumed to be known accurately).
Errors in determining the average velocities $\langle V_a\rangle$ and $\langle V_b\rangle$ were not taken into account when converting the observed frequencies into comoving reference frames, since they are not dominant
in the total error budget $\sigma_{\Delta\mu/\mu}$.

Another possible source of error is the calibration error $\sigma_{\rm sys} \la 50$ kHz, discussed above.
In our case, not absolute values of frequencies are compared, but their differences. Therefore
the calibration error is leveled because the frequencies fall within the same band IF = 4 GHz.

Thus, the result $\langle \Delta\mu/\mu\rangle = (-3.7\pm0.5)\times 10^{-7}$ indicates
the non-zero signal at 7.5$\sigma$.
Even if the frequency differences in these measurements are exposed to systematic shifts of
$\sim$ 50 kHz, the statistical significance of the detected signal is not lost:
$\langle \Delta\mu_{a,b}/\mu \rangle = (-3.7\pm0.7)\times10^{-7} (5.3\sigma$).

If we apply a different estimation method of $\Delta\mu/\mu$ (equation  (\ref{Eq5})) separately to SgrB2(N)
and to Orion-KL, the following mean values are obtained:
 \begin{equation}
\langle{\Delta\mu}/{\mu}\rangle_{a} = (-4.2\pm0.7)\times10^{-7},
\label{Eq12}
\end{equation}
 \begin{equation}
\langle{\Delta\mu}/{\mu}\rangle_b = (-5\pm6)\times10^{-8}.
\label{Eq13}
\end{equation}

These results  show that no signal is detected in Orion-KL,
whereas  SgrB2(N) exhibits a change in $\mu$ at the confidence level of 6$\sigma$.

We also note that the most stringent upper limits on $\Delta\mu/\mu$ (see above) were obtained earlier from observations of molecular clouds in the Galactic disk
located at relatively close distances
to the Solar System, $D_{GC} =$8.5 -- 8.7 kpc
(galactocentric distance of the Solar System $D_{GC} =$ 8.34 kpc \cite{MRZ}).
In this paper, we compare
a relatively close object to the Solar System Orion-KL ($D_{GC}$ = 8.7 kpc), for which the signal at the level of 10$^{-8}$
is not detected, and SgrB2(N) in the center of the Galaxy, showing a lower ratio of $\Delta\mu/\mu$.

Since it is believed that the density ratio of baryonic matter, $\rho_{BM}$,
to dark matter, $\rho_{DM}$, varies along the Galactic radius, and closer to the Galactic center $\rho_{BM}/\rho_{DM} > 1$, but at the periphery
$\rho_{BM}/\rho_{DM} < 1$, it can be assumed that the detected signal in $\Delta\mu/\mu$
correlates with the distribution of dark matter.

In conclusion, we summarize the main results of the work. A statistically significant variation of the fundamental physical
constant $\mu$ at the confidence level of 7.5$\sigma$ was found from a comparison of the same
methanol lines (CH$_3$OH) in two objects spaced approximately 9 kpc apart~-- 
SgrB2(N), located in the center of the Galaxy, and Orion-KL, located closer to the Galactic periphery. 
The measured relative shift of the spectral lines in methanol
is not explained by known systematic effects, and can be interpreted
as a manifestation of the interaction of dark matter with baryonic matter. However, to confirm this
assumption, it is necessary to conduct independent measurements of $\Delta\mu/\mu$
in other sources in the directions of the Galactic center and anticenter.

\subsection*{Funding}
 This work was supported by ongoing institutional funding and was carried out within the framework of the topic of
 the State Assignment of the Ioffe Institute number FFUG-2024-0002. No additional grants to carry out or direct this
 particular research were obtained.

\subsection*{Conflict of interest} The authors of this work declare that they have no conflicts of interest.

\end{document}